\documentclass[twocolumn]{openjournal}
\usepackage{amsmath}
\usepackage{array} 
\usepackage{booktabs}
\usepackage{soul}
\usepackage{longtable} 
\usepackage{multirow}
\usepackage{orcidlink}
\usepackage{hyperref}

\newcommand{\mbh}{$M_{\rm BH}$}
\newcommand{\msig}{$M_{\rm BH} \text{--} \sigma$}
\newcommand{\sigmasign}{$\sigma$}
\newcommand{\mhalo}{$M_{\rm Halo}$}
\newcommand{\mstar}{$M_{\star}$}
\newcommand{\sersic}{$\acute{n}$}

\begin{document}

\title{Black Hole-Host Galaxy Correlations with Machine Learning: A Comparative Study of Illustris, TNG, and EAGLE}

\author{Jacob Reinheimer $^{1}$}
\author{Yuan Li \,\orcidlink{0000-0001-5262-6150}$^{\ast, 2}$}
\author{Trung Ha \,\orcidlink{0000-0001-6600-2517}$^{2}$}
\author{Melanie Habouzit $^{3,4}$}
\author{Brandon M. Matthews $^{1}$}
\author{George Blaney $^{1}$}
\affiliation{$^1$Department of Physics, University of North Texas, Denton, TX 76203, USA}
\affiliation{$^2$Department of Astronomy, University of Massachusetts Amherst, Amherst, MA 01003, USA} 
\affiliation{$^3$Department of Astronomy, University of Geneva, Chemin d'Ecogia, CH-1290 Versoix, Switzerland}
\affiliation{$^4$Max-Planck-Institut f\"ur Astronomie, K\"onigstuhl 17, D-69117 Heidelberg, Germany}

\email[$^\ast$]{yuanli@umass.edu}

\begin{abstract}
Supermassive black holes (SMBHs) are known to correlate with many properties of their host galaxies, but we do not fully understand these correlations. The strengths (tightness) of these correlations have also been widely debated. In this work, we explore SMBH-host relations in three state-of-the-art cosmological simulations: Illustris, TNG, and EAGLE. Using a variety of machine learning regressors, we measure the scaling relations between black hole mass ($M_{\rm BH}$) and galaxy properties including stellar velocity dispersion ($\sigma$), stellar mass ($M_{\star}$), dark matter halo mass ($M_{\rm Halo}$), and the Sersic index. We find that machine learning regressors provide predictive capabilities superior to linear regression in many scaling relations in simulations, and Multi-layer Perceptron (MLP) regressor has the strongest performance. SMBH-host relations have different strengths in different simulations as a result of their sub-grid models. Similar to the observations, the $M_{\rm BH} $-$\sigma$ relation is a strong correlation in all simulations, but in TNG, the $M_{\rm BH} $-$M_{\star}$ relation is even tighter than $M_{\rm BH} $-$\sigma$. EAGLE produces the weakest SMBH-host correlations among all simulations. Low mass SMBHs tend to be poorly correlated with their host galaxies, but including them can still help machines better grasp the correlations in Illustris and TNG. Combining galaxy properties that strongly correlate with $M_{\rm BH} $ but poorly correlate with each other can improve MLP's performance. $M_{\rm BH} $ is most accurately predicted when all galaxy properties are included in the training, suggesting that SMBH-host correlations are fundamentally multi-dimensional in these simulations.

\end{abstract}

\section{Introduction} \label{sec:intro}

Supermassive black holes (SMBHs) with masses ranging from millions to billions of $M_{\odot}$ are found at the center of nearly all massive galaxies \citep{Kormendy_1995,Irina_2020,Ma_2014,Magorrian_1998}. These objects and their relation to their host galaxies have been focal points of astrophysical research.

Observations suggest that SMBH feedback can strongly impact their host galaxies. During periods of active accretion, SMBHs release significant energy and momentum, which can drive jets and/or massive outflows that heat the interstellar and circum-galactic medium, thereby regulating star formation and affecting the overall evolution of their host galaxies. In the local universe, X-ray cavities driven by the radio-mode jet feedback from the center of galaxy clusters have been directly observed \citep{McNamara_2007, Fabian_2012}. At higher redshifts, the detection of ultra-fast outflows in quasar spectra provides evidence of SMBH feedback at high accretion rates \citep{Tombesi_2015, King_2015}. Additionally, the impact of these outflows on the molecular gas content—a critical component for star formation—has been extensively studied, suggesting a direct link between SMBH activity and the suppression of star formation in quasar host galaxies \citep{Feruglio_2015, Cicone_2014}. These phenomena support some form of co-evolution between SMBHs and their host galaxies, where the energy output from SMBHs during active phases is crucial in sculpting galaxy properties \citep{Kormendy_2013, Harrison_2018}.

The co-evolution of SMBHs and their host galaxies can also be studied through correlations between SMBH mass (\mbh) and various host galaxy properties. 
Early works identified a relation between \mbh\ and the galaxy's bulge luminosity, as well as its mass \citep{Kormendy_1993,Magorrian_1998,Gultekin_2009,Graham_2001}. \citet{Gebhardt_2000} and \citet{Merritt_2001} identified another strong correlation between \mbh\ and the stellar velocity dispersion, \sigmasign, of their host galaxies. The \msig\ relation has since been refined through advances in observational techniques, such as adaptive optics and integral field spectroscopy \citep{Cappellari_2016}, and these technological improvements have helped expand our understanding of the relation \citep{McConnell_2013}. 

Further exploration of SMBH-host relations reveals that \mbh\ also correlates with the total stellar mass \citep[\mstar;][]{Reines_2015,Terrazas_2016,Schulze_2011}, as well as with the dark matter halo mass \citep[\mhalo;][]{Ferrarese_2002,Booth_2011,Bandara_2009}. \mbh\ has also been found to correlate with structural properties of the host galaxies, such as the Sersic index (\sersic), which is related to the compactness of their corresponding host galaxy \citep[\sersic;][]{Savorgnan_2016, Sersic_1968}, and the pitch angle of the spiral arms in disk galaxies \citep{Seigar_2008}.
This indicates a complex, multi-dimensional dependency between black hole growth and host galaxy properties. 

The physical interpretations of the \msig\ relation remain an active area of research.
While SMBH-host relations may be related to SMBH feedback, other theories suggest that some of them can also arise from the central limit theorem \citep[e.g.,][]{Hirshmann_2010,Jahnke_2011}. 
Additionally, it has also been suggested that these scaling relations appear even in numerical simulations without AGN Feedback \citep{Angles_2013}. More work is needed to understand the complexities behind these relations.

The detailed mechanisms of SMBH feedback are also complicated, but recent advancements in general-relativistic magnetohydrodynamic simulations have significantly deepened our understanding of both the dynamics in accretion disks and the initialization of jets and winds from active galactic nuclei (AGN). Studies like those by \citet{Fragile_2007,Tchekhovskoy_2011,Bu_2016,Jiang_2019} have shed light on these small-scale processes. At larger scales, simulations that span galactic to cosmological distances have been instrumental in testing various AGN feedback models, revealing the diverse impacts these mechanisms have on galactic environments, star formation processes, and overall black hole activity. This broad spectrum of research has enriched our understanding of how black holes influence their host galaxies \citep{Omma_2004,Guo_2011,Choi_2012,Li_2015,Yang_2016,Qiu_2019,Yuan_2018}.

On larger scales, cosmological simulations like Illustris \citep{Vogelsberger_2014}, TNG \citep{Springel_2017,Pillepich_2017,Marinacci_2018,Nelson_2017,Naiman_2018}, EAGLE \citep{Schaye_2014,Crain_2015}, Horizon-AGN \citep{Dubois2014, Dubois_2016}, SIMBA \citep{Dave_2019}, and many others, incorporate different subgrid models of SMBH seeding, feeding, and feedback mechanisms. 
These models offer important insights into the dynamical processes of galaxy formation and evolution that remain elusive in direct observations. These cosmological models have served as excellent labs to study the SMBH-host coevolution. For example, \citet{Li_2020,Terrazas_2020,Habouzit2022,Voit_2024, 2025MNRAS.543.2489Z} have studied SMBH-host galaxy scaling relations across different cosmological simulations, and have revealed how these relations can be shaped by the different feedback subgrid models.

Alongside advances in computational simulations, machine learning has recently emerged as a powerful tool for probing complex phenomena in astrophysics \citep[e.g.,][]{Olney2020, Xu2020}. Machine learning algorithms can uncover patterns and correlations that may be too subtle or complex for traditional analytical methods. These techniques are particularly adept at handling the high-dimensional hypterparameter spaces and non-linear interactions inherent in astrophysical data \citep[e.g.,][]{ha_2024}. In recent years, many works have used machine learning to quantify correlations related to galaxy evolution and identify the relative importance of different features in multi-dimensional relations \citep[e.g.,][]{Ellison_2020, baker_2023,Davis_2023,Bluck_2022,Bluck_2023,Thongkonsing_2023, Jin_2024}.

In this work, we study SMBH--host galaxy relations at $z=0$ in three state-of-the-art cosmological simulations: Illustris, TNG, and EAGLE \citep{Schaye_2014,Crain_2015}. We use a variety of machine-learning regression methods to measure the relative tightness of each relation, as well as the correlation between SMBH and the combination of multiple galaxy properties. The structure of the paper is as follows: Section 2 describes the cosmological simulations and the galaxy properties used in our analysis. Section 3 describes our methodology and the machine learning algorithms employed, as well as the hyperparameter optimization process for achieving the best-fitting models. In section 4.1, we present all SMBH-host scaling relations analyzed in this work. In 4.2, we quantify the tightness of these relations using machine learning regressors. Section 4.3 examines how these relations depends on \mbh\. Section 4.4 discusses higher-dimensional correlations between \mbh and galaxy properties. We summarize our work in Section 5.

\section{Data} \label{sec:sims}

\subsection{Simulations}

We obtain simulation snapshots from the Illustris \citep{Vogelsberger_2014}, TNG \citep{Springel_2017,Pillepich_2017,Marinacci_2018,Nelson_2017,Naiman_2018}, and EAGLE \citep{Schaye_2014,Crain_2015} simulations. 
These are large-scale cosmological simulations that model gas cooling, star formation, stellar feedback, and black hole feeding and feedback, among other physical processes related to galaxy evolution.
All simulations have a comoving box size of $100^3$ cMpc$^3$. 
Below, we describe the key aspects of each simulations and their most relevant subgrid treatments to model SMBHs, such as the SMBH seed mass, accretion, and feedback (see Table~\ref{tab:sims} for a summary).

\subsubsection{Illustris}

Illustris is an N-body/hydrodynamical cosmological simulation with the moving mesh code \texttt{AREPO}. It solves second-order ideal hydrodynamic equations using a Godunov-type scheme with an exact Riemann solver. Illustris uses relatively low SMBH seed mass of $1.42 \times 10^5 M_{\odot}$ \citep{Sijacki2015}. To model BH accretion, Illustris uses a boosted Bondi accretion rate of $\dot{M}_{\rm BH} = \alpha \dot{M}_{\rm Bondi}$, where $\alpha = 100$ to account for the unresolved Bondi radii. Based on the Eddington ratio, it uses two different modes to model SMBH feedback with the energy output of $\dot{E}_{\rm out}=\epsilon \dot{M}_{\rm BH} c^2$. The first mode, called the low- or radio-mode, has a higher feedback efficiency, $\epsilon = 0.35$. In this mode, feedback is injected back into the system via randomly placed $\sim50$ kpc thermal bubbles displaced away from the central galaxy. The second, called the high- or quasar-mode, operates with a lower efficiency of $\epsilon = 0.05$. In this mode, feedback is implemented as thermal energy injection near the BH.

\subsubsection{TNG}

TNG (``The Next Generation") is a successor to Illustris. It includes magnetohydrodynamics with the same moving mesh code \texttt{AREPO} \citep{Weinberger2020}. TNG's \mbh\ seed mass is $1.18 \times 10^6 M_{\odot}$. It is higher than Illustris seed mass and is also the highest among all three sets of simulations analyzed here. It uses an un-boosted Bondi rate for black hole accretion. The high accretion mode feedback scheme is the same as Illustris, but the low accretion mode is different, where TNG injects a kinetic wind in a random direction, rather than the thermal bubbles. This change is meant to make SMBH feedback more efficient. There is also a difference in the critical Eddington ratio used to separate the modes. In TNG, this critical ratio varies with the mass of the black hole, making the more massive ones able to enter the low mode more easily \citep{Weinberger2018}. 

\subsubsection{EAGLE}

EAGLE (``Evolution and Assembly of GaLaxies and their Environments") is another large-scale cosmological simulation. It uses a modified N-Body Tree-PM smoothed particle hydrodynamics (SPH) code \texttt{GADGET 2} \citep{Springel_2005} with subgrid physics based on \texttt{OWLS} \citep{Schaye_2010} and \texttt{COSMO-OWLS} \citep{Le_Brun_2014}. We use the L0100N1504 simulation of EAGLE which has the same comoving size as Illustris and TNG. EAGLE's BH seed mass is similar to Illustris, at $1.475 \times 10^5 M_{\odot}$. EAGLE employs a different feedback scheme from Illustris and TNG. It only models one mode that uses stochastic thermal energy injection, where $\epsilon = 0.015$, which is near the middle value of Illustris and TNG. This energy is only released when enough energy has been accumulated to raise the temperature above $10^{8.5}K$.

\begin{deluxetable*}{cccc} 
\tablecaption{Simulation Summary\label{tab:sims}}
\tablehead{\colhead{Simulation} & \colhead{BH seed mass} & \colhead{Accretion} & \colhead{Feedback modes}}
\startdata
Illustris & $1.42 \times 10^5 M_{\odot}$ & Boosted Bondi & High (thermal) and low ($\sim 50$\ kpc thermal bubbles)\\
TNG & $1.18 \times 10^6 M_{\odot}$ & Bondi & High (thermal) and low (kinetic wind in random directions)\\
EAGLE & $1.475 \times 10^5 M_{\odot}$ & Bondi-Hoyle-Lyttleton &  Thermal energy injection if $T > 10^{8.5} K$\\
\enddata
\end{deluxetable*}

\subsection{Data Selection} \label{sec:data_selection}
We extract the following parameters from each simulation for our analysis: $M_{BH}$, $\sigma$, \mstar, \mhalo, and \sersic. We discuss how these parameters are defined and obtained from each simulation in this subsection. 

\begin{figure}[ht]
\centering
\includegraphics[scale=.35]{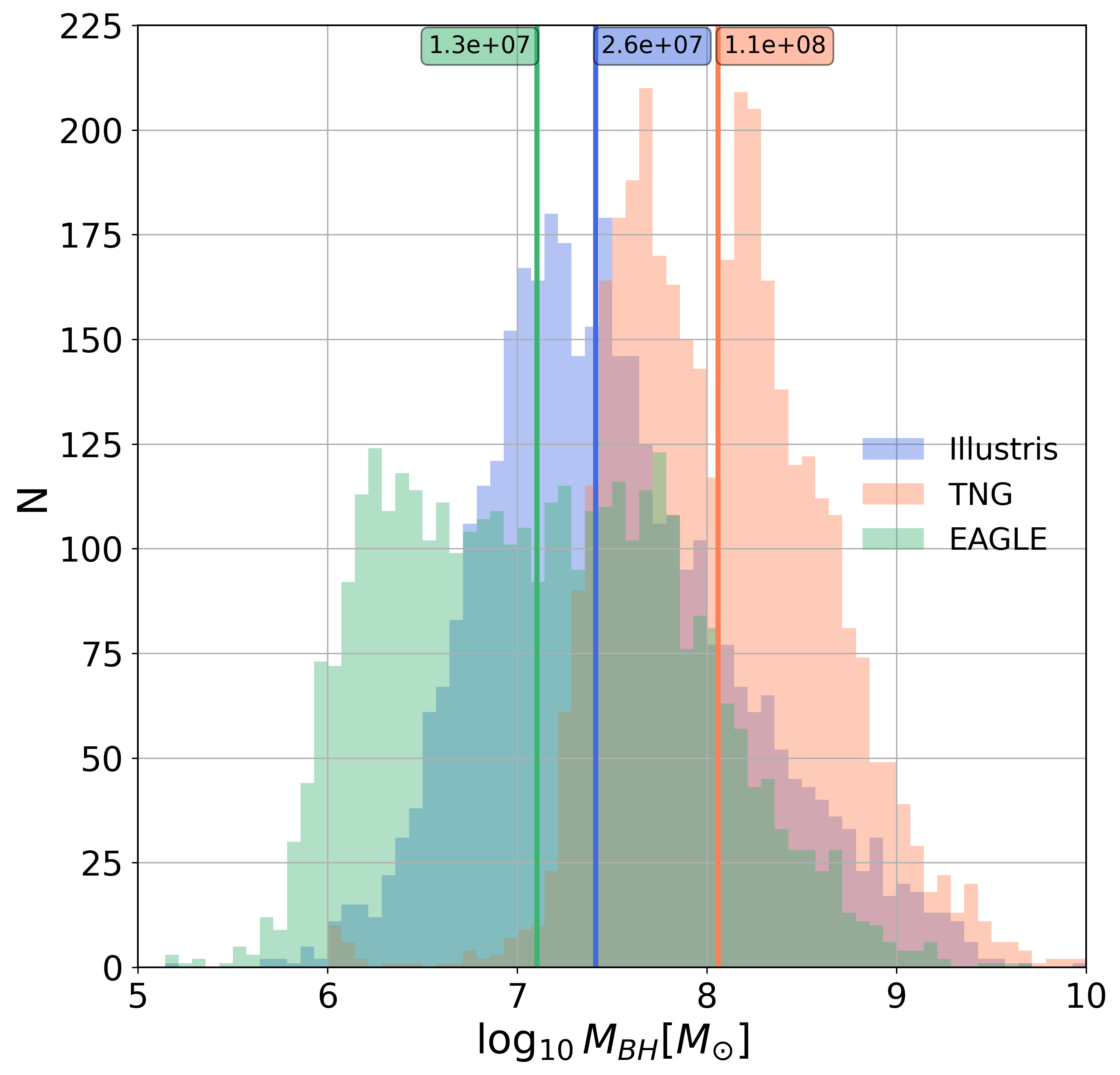}
\caption{\mbh\ distributions at $z=0$ for the Illustris (Blue), TNG (Orange), and EAGLE (Green) simulations. Each vertical line represents the median value of the \mbh\ in that simulation sample (the most massive 3607 galaxies), and is used to divide each sample into high- and low-mass SMBHs for the analyses in Section~\ref{sec:mass_dependence}.}
\label{fig:hist}
\end{figure}

The \mbh\ in Illustris and TNG is defined as the sum of all masses of all the black hole particles in the halo. In practice, this is the mass of the central black hole for most galaxies, as only a few of the most massive galaxies host more than one black hole particle. For EAGLE, we define a spherical aperture of size 100 kpc, then \mbh\ is the sum of all black hole particles within the aperture. The distributions of $M_{BH}$ in different simulations are shown in Figure~\ref{fig:hist}.

$\sigma$ for Illustris and TNG is the rest-frame SDSS r-band luminosity-weighted stellar line-of-sight velocity dispersion measured within a projected radius of 1.5" from the galaxy center in the x projection \citep{Xu_2017}. In EAGLE, it is the one-dimensional velocity dispersion of stars within the aperture, calculated as $(\frac{2 E_k}{3 M_{\rm star}})^{1/2}$ where $E_k$ is the stellar kinetic energy, and $M_{\rm star}$ is the stellar mass.

The stellar mass \mstar\ in Illustris and TNG is the sum of all particle masses within twice the stellar half-mass-radius. In EAGLE, it is the total stellar mass within the 100 kpc aperture.

\mhalo\ in Illustris and TNG is defined as the sum of all particles within the halo. For EAGLE, we first obtain $M_{\rm DM}$ for dark matter particles within the aperture. In order to have similar comparisons, we then add all masses together (including $M_{BH}$, \mstar, and gas mass) to make EAGLE have a similar parameter for comparison. 

\sersic\ describes the compactness of a galaxy based on its projected profile. We use \sersic\ obtained via synthetic observations. For Illustris and TNG, synthetic images were designed to match Pan-STARRS observations \citep{Rodriguez-Gomez_2019}. For EAGLE, \citet{De_Graaff_2022} provides \sersic\ using synthetic images of simulated galaxies created with the SKIRT \citep{Baes_2011} radiative transfer code. \citet{De_Graaff_2022} provides \sersic\ values for only those galaxies with \hbox{$M_{\star} \gtrsim 10^{10} M_{\odot}$}, resulting in a sample size of 3607 candidate galaxies. This limits our investigation to the 3607 most massive (\mstar) galaxies in each simulation. We use the same number of galaxies from each simulation to ensure a fair comparison. 

\section{Methods} \label{sec:ml}

\subsection{Measuring Black Hole Scaling Relations using Machine Learning}
Our study employs a variety of machine learning algorithms (regressors), each using distinct fitting techniques, which we describe in more detail in the following subsections. We apply these regressors to measure the strength of the previously mentioned SMBH--host galaxy scaling relations. 
We first log-transform (base 10) and normalize all data to scale the maximum and minimum values to one and zero, respectively. Next, we randomly split the data into training (70\%) and testing (30\%) sets, with the former used for model training and the latter for performance evaluation. 
We feed the training data to our regressors and use the best-fit model to predict $M_{BH}$ for the testing set for a given galaxy property or a combination of properties. 
We compute the mean squared error (MSE) that serves as a one-to-one comparison between different regressors, as well as different galaxy properties. The MSE measures how well predicted quantities (\mbh in our study) compares with the true values, and is calculated using the formula $\frac{1}{N} \sum_{i=0}^{N-1}(y_{i}-\hat{y}_{i})^2$.

\subsection{Machine Learning Regressors}

We employ five widely-used machine learning algorithms from the Python package \texttt{Scikit-learn} \citep{pedregosa2018scikitlearn}: linear regression, decision tree, random forest, and multi-layer perceptron (MLP). 
Additionally, we use XGBoost from \citet{xgboost}.

\subsubsection{Linear Regression}
Linear regression models the relationship between a dependent variable and one or more independent variables by fitting a linear equation to the data. It aims to minimize the difference between actual and predicted values using the least squares method. Its simplicity and interpretability make it widely used in both predictive modeling and inferential analysis.

\subsubsection{Decision Tree}
A decision tree is a supervised learning algorithm used for classification and regression tasks. It works by recursively splitting the dataset into subsets based on specific decision criteria, often focusing on maximizing homogeneity within the resulting groups. Each internal node represents a feature, each branch represents a decision rule, and each leaf node represents an outcome or class. The algorithm aims to find splits that best separate the data, using the MSE for such decisions.

\subsubsection{Random Forest}

A random forest regressor is an ensemble learning method that combines multiple decision trees to enhance the accuracy and reduce overfitting. It operates by creating a "forest" of random decision trees, each trained on a random subset of the data and using a subset of features at each split. This diversity among trees increases the model's robustness and generalization capability. Predictions are made by averaging the outputs of all trees, which reduces variance and improves performance on unseen data, making random forest a powerful and versatile machine learning technique.

\subsubsection{XGBoost}

The XGBoost regressor, akin to a random forest, is an ensemble method that builds decision trees sequentially, each one enhancing the accuracy by correcting errors from its predecessors—a process known as 'boosting.' A key feature of XGBoost is the use of regularization, which reduces model complexity and helps prevent overfitting by penalizing larger values in the model's learning function. This ensures more robust performance on out-of-distribution data, which refers to new data that differs in some way from the data used in training.

\subsubsection{Multi-layer Perceptron}

MLP is a type of neural network known for its ability to model complex, non-linear relationships between inputs and outputs. 
Unlike a single-layer perceptron that can only handle linear separations, the MLP uses multiple layers of nodes, or neurons, each applying a non-linear activation function. 
This architecture allows the MLP to learn deep patterns in the data through backpropagation, optimizing weights across layers to minimize prediction errors. 
It is particularly effective in scenarios requiring the capture of intricate patterns in high-dimensional data.

\subsection{Hyperparameter Optimization}\label{sec:param_search}

Each machine learning regressor has a set of adjustable hyperparameters, such as the initial learning rate, alpha, hidden layer size, and the number of hidden layers. These hyperparameters impact how the regressor conducts its fit, allowing for the fine-tuning of its performance. 
For instance, in the decision tree regressor, the "max depth" hyperparameter determines the maximum height of a tree. 
Adjusting these hyperparameters can lead to varying complexity of the model and fit qualities. 
In Section~\ref{sec:mse}, we conduct a hyperparameter search to find the best-performing set of hyperparameters for each model. We use the sklearn RandomizedSearchCV (cross-validated) function. It implements a randomized grid search technique, where a subset of random hyperparameter combinations is sampled (as opposed to GridSearchCV, which tries out all parameter values) to identify the optimal configuration that produces the lowest MSE for each regressor. This hyperparameter optimization exercise ensures a fair comparison between different regressors at a reasonable computational cost. More details of the hyperparameters are listed in the Appendix.

\section{Results and Discussion} \label{sec:corrs}

\subsection{SMBH--Host Galaxy Scaling Relations in Cosmological Simulations}

\begin{figure*}[hbt!]
\centering
\includegraphics[width=\linewidth]{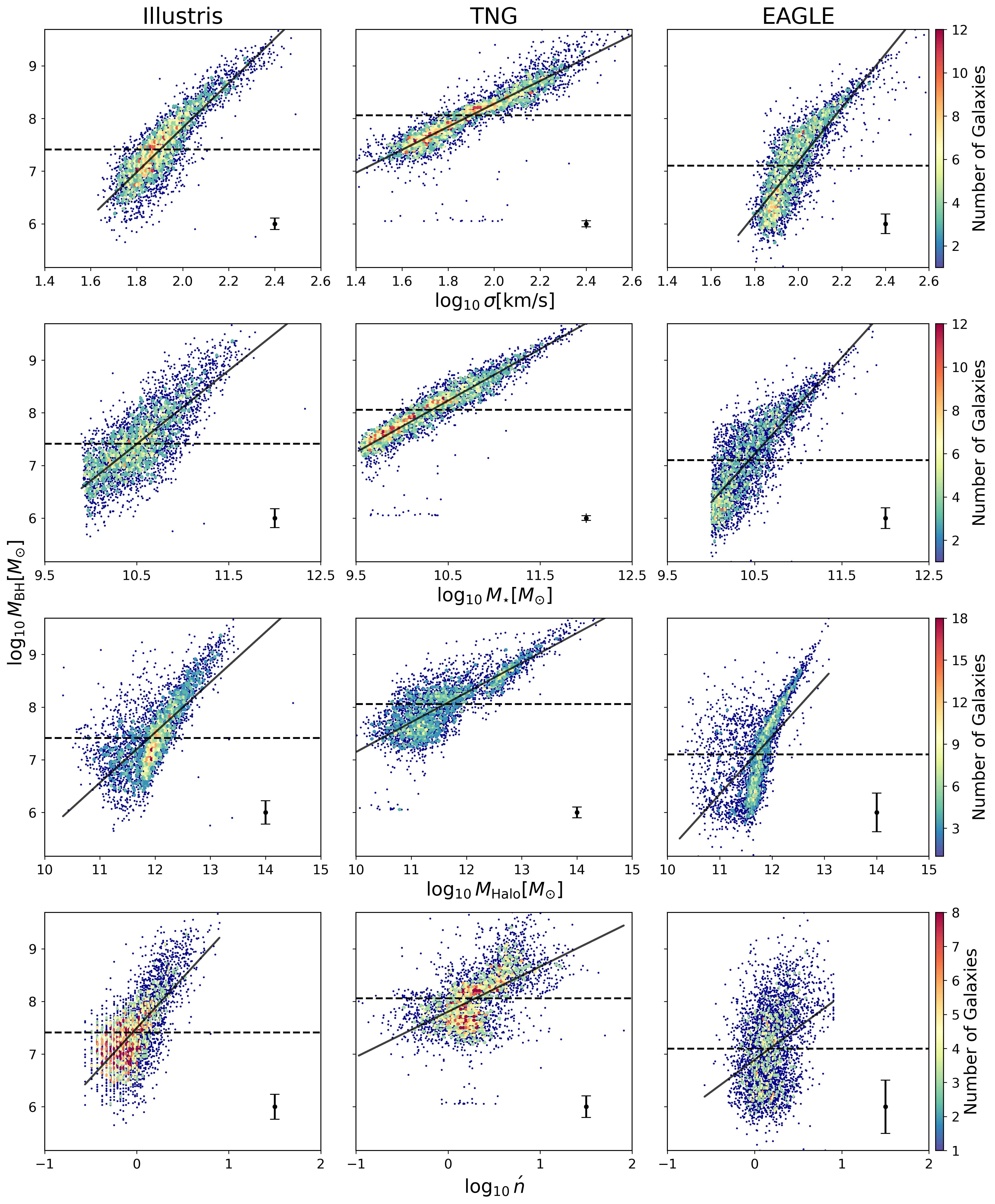}\caption{SMBH--host galaxy correlations at $z=0$ in the Illustris (left), TNG (middle), and EAGLE (right) simulations. From top to bottom, we show the \msig\, the \hbox{\mbh--\mstar}, the \hbox{\mbh--\mhalo}, and the \hbox{\mbh--\sersic} relations. Only the most massive 3607 galaxies in each simulation are included in our study, which corresponds to $M_{\star} \gtrsim 10^{10} M_{\odot}$ for EAGLE. The black solid line is a linear least squares regression fit, and the error bar in the bottom right describes one standard deviation errors.}
\label{fig:scale_rels}
\end{figure*} 

Figure~\ref{fig:scale_rels} shows the scaling relations between SMBHs and their host galaxy properties ($\sigma$, \mstar, \mhalo, and \sersic) that are studied in this work, with data from Illustris, TNG, and EAGLE. 
Across all three simulations, a consistent monotonic relationship exists between \mbh\ and the majority of host galaxy properties analyzed. 
This is in general agreement with the observed trends for these scaling relations \citep{Kormendy_1993,Magorrian_1998,Ma_2014,Graham_2016}. In this work, we focus on the analysis and comparison among simulation results. We leave the comparison with the observational data for future work. 

For each SMBH--host galaxy scaling relation, each simulation produces different amplitudes, slopes, scatters, and sometimes different shapes. 
The black line atop each plot represents the ordinary least square (OLS) linear fit, and the accompanying black error bars illustrate the standard error. 
A large error of a certain relation can stem from a large scatter in the data or a large deviation from a simple linear shape. 
For example, \mbh--\sersic\, tends to have a large scatter in all simulations. In Illustris's \mbh--\mhalo\ relation, the notable bulge-like phenomenon creates a non-linear feature that skews the linear fit away from the population's observed center. 
Several relations have a tear-drop shape with a higher scatter at lower \mbh\ values, a phenomenon we explore further in Section~\ref{sec:mass_dependence}. 

While linear regression adequately describe linear relationships, such as TNG's \mbh--\mstar, this method struggles with more non-linear correlations. In the following sections, we incorporate machine learning and include various algorithms with non-linear capabilities and more flexibility.

\subsection{Measuring SMBH--Host Galaxy Relations Using Machine Learning} \label{sec:mse}

\begin{figure*}
\centering
\includegraphics[width=0.99\textwidth,trim=0cm 2.6cm 0cm 0cm, clip=true]{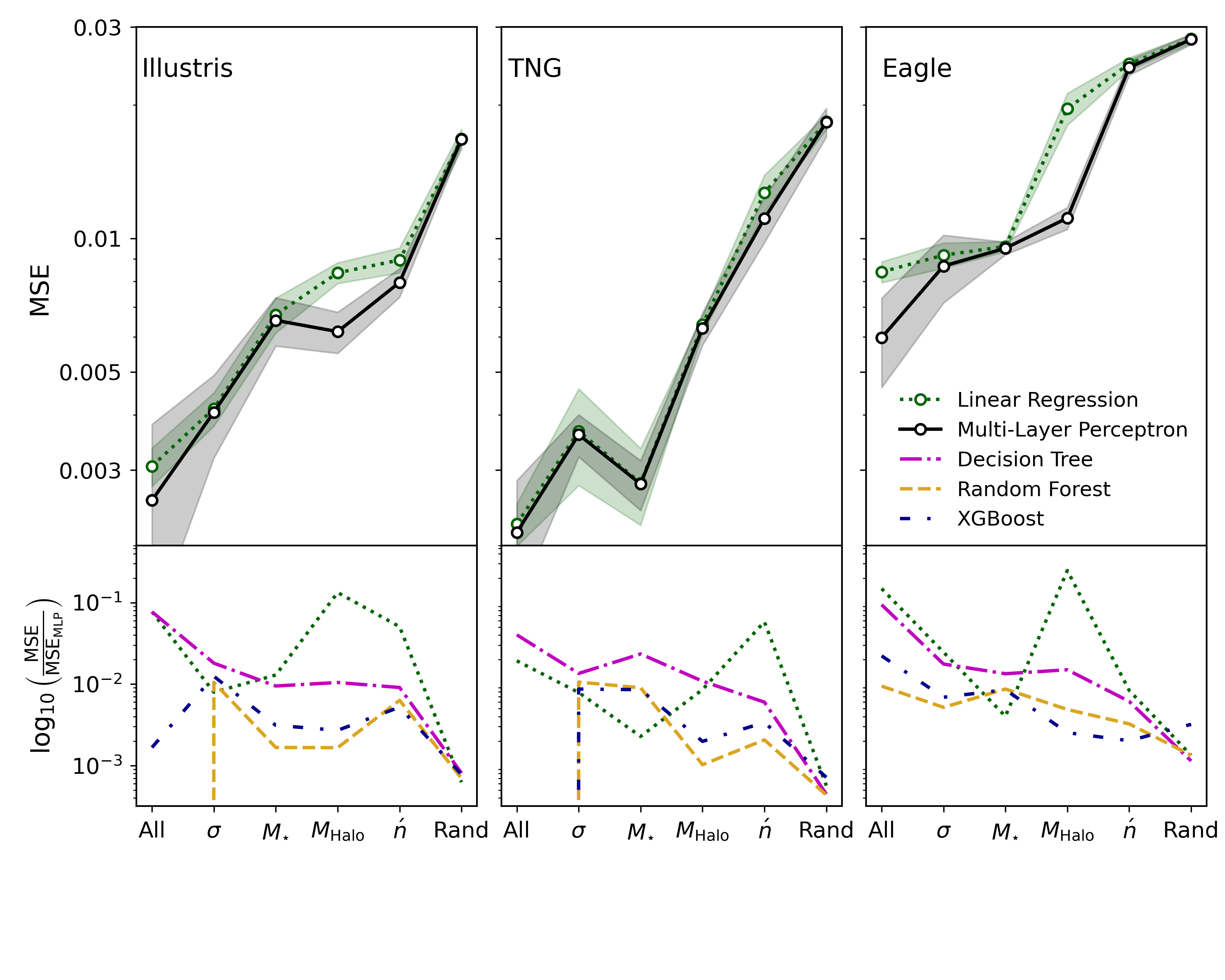}\caption{The strength of SMBH-host correlations in Illustris (left), TNG (middle), and EAGLE (right) measured by machine learning regressors' ability to predict \mbh. Small MSEs (errors in predicted \mbh) correspond to tight correlations.
We repeat the analysis 50 times for each regressor to quantify the uncertainty from the random training/testing split. Solid lines in the top panels connect the average MSEs of the 50 realizations while the shaded regions show the standard deviation. For clarity, we only show linear regressor and MLP in the top panels. In the bottom panels, the other regressors's performances are divided by MLP to show their relative strengths.}
\label{fig:MSE_results}
\end{figure*}

We use machine learning regressors and measure their resulting MSE for different SMBH-host relations to quantify how well they predict \mbh, which indicates how tight the relations are. For each regressor, we perform a broad hyperparameter search as is described in section~\ref{sec:param_search}. The top panels of Figure~\ref{fig:MSE_results} show the resulting MSEs for linear regression and the MLP regressor. The bottom panels show the comparison between other regressors normalized by MLP. We use the linear regressor from sklearn following the same procedure as the other machine learning regressors for consistency, but the coefficients of the machine learning linear regressor we obtain are identical to the least square linear fit in Figure~\ref{fig:scale_rels}. 

Figure~\ref{fig:MSE_results} shows the linear regressor's strong performance on scaling relationships that are shaped linearly, such as TNG's \mbh--\mstar, but also highlights its struggles with non-linear relationships, such as \mbh--\mhalo\ in Illustris and EAGLE. 
For these intrinsically non-linear relations, the linear regressor would under-perform compared to the other machine learning regressors, because they are more complex and adaptable. 
Even for the seemingly linear relations, MLP performs slightly better than linear regressor (see the bottom panels of Figure~\ref{fig:MSE_results}). 

In the bottom panels of Figure~\ref{fig:MSE_results}, the performance metrics of the other regressors are normalized by the outcomes of the MLP regressor. 
When the relations are individually input into the regressors, the MLP regressor consistently outperforms all other regressors. 
This is because neural networks such as the MLP regressor are superior at capturing non-linear correlation through their layered structure and the use of (often non-linear) activation functions.
When trained on a single relationship like \msig, they are particularly effective at modeling the specific nonlinear dynamics of that relationship. 
Out of the remaining machine learning regressors, random forest and XGBoost perform better than decision tree. 
Since they are both statistical ensembles of the decision tree regressor (see Section~\ref{sec:ml} for details), their performance is superior to decision tree, as one would expect. They also outperform linear regression in most cases, while decision tree falls behind linear regression for many relations.

One of the categories in the x-axis of Figure~\ref{fig:MSE_results} is ``All". This relation describes a multi-dimensional correlation in which all four host galaxy properties (\mstar, \mhalo, \sigmasign, and \sersic) are considered simultaneously.
Such a relation performs markedly better than single-property relations due to increased dimensionality of the learned parameters. We explore higher dimensional SMBH-host correlations in more detail in Section~\ref{sec:multi}.

When we train on all relations, MLP continues to exhibit great performance, particularly in EAGLE. 
However, for the Illustris and TNG datasets, XGBoost and random forest demonstrate superior performance to MLP. 
This is because when all relations are included, the data complexity increases. 
The additional complexity means models with robust feature handling and generalization capabilities, like XGBoost and random forest, are likely better suited to adapt and learn from this increased complexity. They can dissect the data into more manageable pieces (trees), each focusing on different aspects of the data, thereby achieving a more comprehensive understanding. Additionally, they benefit from their ensemble learning techniques, which allows better generalization and less risk of harmful overfitting. 

We have also created an array of normalized random values between 0 and 1 as a test. The models also train on this random array to predict \mbh, which is shown as the last element in each panel of Figure \ref{fig:MSE_results}. This establishes a baseline performance for each model and is also an indirect measurement of the \mbh spread in each simulation, as we discuss later.

Overall, the MLP regressor, XGBoost, and random forest show similarly strong performances, with MLP showing the smallest MSEs for most of the relations. 
In our subsequent analysis, we use only the MLP regressor as a representative machine learning model for simplicity. 
For each BH--host galaxy relation, we perform an extensive hyperparameter search with MLP to optimize its performance, as described in Section~\ref{sec:ml}. 
These models form the foundation of our analysis in Sections 4.2 and 4.3. 

Within the Illustris simulation, the \msig\ relation shows the smallest MSE when compared to the other individual relations, independent of the regressors used. Such a strong correlation appears in agreement with the observations \citep{Tremaine_2002, Gultekin_2009, Kormendy_2013}. In particular, \citet{Shankar2025} performed both pairwise residual analysis and machine learning on a sample of local galaxies. They identified \sigmasign\ as the galaxy property most correlated with \mbh. The MSEs of the \mbh--\mstar, --\mhalo, and --\sersic\ relations are comparable in Illustris when measured with MLP. Although the linear regressor performs well for a few of these relations, it is unable to explain for the relationship between, e.g., \mbh\ and \mhalo.

In the TNG simulation, the \msig\ relation is also strong, showing a similarly small MSE as Illustris' \msig\ relation. However, unlike Illustris and EAGLE, \msig\ is not the strongest relationship in TNG. Instead, the \mbh--\mstar\ relation is even stronger, with the smallest MSE among all the individual relationships studied in this work. This very tight \mbh--\mstar\ correlation in TNG has also been reported in many previous studies \citep{Weinberger2018, Habouzit2021, 2025MNRAS.543.2489Z}. \citet{Li_2020} speculates that the tightness has to do with the accretion scheme in TNG, where SMBH accretion rate is computed as a kernel-weighted average over 256 neighboring cells. Given their typical spatial resolution of the star-forming ISM gas ($\sim 100-140$ pc), the SMBH ``accretion zone'' has an effective size of $\sim 1$ kpc. This likely helps synchronize the growth of \mbh\ and \mstar\ much better than simulations with SMBH accretion schemes that only use very local cells. 


In the EAGLE simulation, \msig\ is the best-performing relation, but \mbh--\mstar\ and --\mhalo\ both have relatively similar performance. 
Like Illustris, we see the same gap between the linear and MLP regressors in \mhalo, as the linear regressor cannot adequately describe the relation. 
When EAGLE is trained on the \sersic, the resulting MSE is similar to the MSE when trained on random data points, which implies that the \sersic\ values reported in EAGLE do not have a statistically significant correlation with the BH evolution.


EAGLE reports the highest overall MSE across the three simulations, an indication that the SMBHs are not well-coupled to the properties of their host galaxies. 
This may be partly attributed to the simulation's BH feedback model, where energy is not instantaneously released. 
Instead, it is re-inputted into the local environment if the surrounding temperature reaches a critical threshold of $10^{8.5}K$. 
This delayed feedback model may effectively weaken the intrinsic BH--host galaxy relations. 
A wider BH mass distribution at a fixed sample size may be another factor that contributes to the larger scatter in the EAGLE BH--host galaxy relations. The large BH mass scatter has to do with the difficult growth of BHs in low-mass galaxies in Eagle as a result of efficient SN feedback \citep{Habouzit2021}.
As Figure~\ref{fig:hist} shows, EAGLEs \mbh\ distribution is broader than both Illustris and TNG, making it harder for machines to predict the correct \mbh. 
The last element in each panel of Figure~\ref{fig:MSE_results} is an array of random numbers which have no intrinsic correlation with \mbh. 
Illustris and TNG's regressors have a similar performance when trained on random values. 
However, EAGLE's random value regressors have a noticeably higher MSE. This is the result of EAGLE's broader black hole mass distribution. 

Another aspect of the feedback model that will affect these relations is the feedback efficiency $\epsilon$. As discussed in Section~\ref{sec:data_selection}, Illustris and TNG use the same feedback scheme where $\epsilon = 0.35$ in radio-mode and $\epsilon = 0.05$ in quasar-mode. On the other hand, EAGLE only has a single mode of feedback where $\epsilon = 0.15$. 
This value is lower than Illustris and TNG, potentially resulting in a weaker effect of EAGLE's SMBH feedback on its host galaxy's properties. 
This could be another reason why our machine learning regressors fit more poorly on EAGLE, as the SMBHs are less correlated with their hosts than Illustris and TNG.

While the linear regressor is an effective approximator of many BH--host galaxy relationships, the MLP regressor captures the intrinsic non-linearities of the data more effectively.
After extensively optimizing its hyperparameters, we achieve the lowest MSE regressor for each relation and simulation. 
We additionally observe the \msig\ relationship being the best predictor of \mbh\ in Illustris and EAGLE, but \mstar\ is the best in TNG. 
Additionally, we find that TNG has overall the lowest MSE, while EAGLE has the largest MSE. 
These are likely a result of different subgrid models in the simulations, especially the ones related to SMBHs. 
To further analyze this phenomenon, we explore how the SMBH-host relations depend on \mbh\ in the next subsection. 

\subsection{The \mbh\ Dependence} \label{sec:mass_dependence}

\begin{figure*}[ht]
\centering
\includegraphics[width=0.99\textwidth]{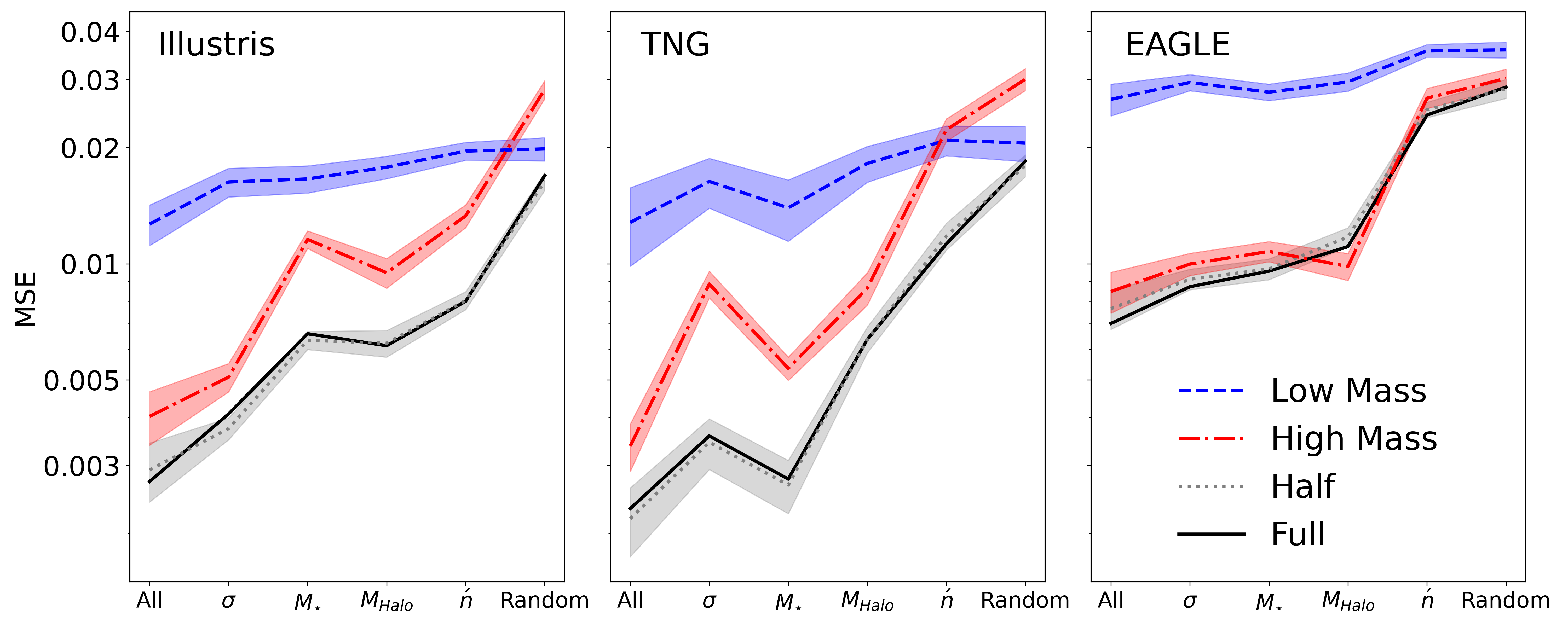}\caption{The tightness of SMBH-host relations measured with optimized MLP regressor for low- and high-mass SMBHs in Illustris (left), TNG (middle), and EAGLE (right). Low- and high-mass SMBHs are defined as SMBHs below and above the median \mbh\, in each sample (see Figure~\ref{fig:hist}). Similar to Figure~\ref{fig:MSE_results}, the lines connect the average MSEs of 50 iterations and the shaded regions represent the standard deviation. For comparison, we also show the full sample (same as the black line in Figure~\ref{fig:MSE_results}) and a randomly selected half sample (see Section~\ref{sec:mass_dependence} for details).}
\label{fig:MBH_dependence}
\end{figure*}

We examine how the strength of the BH--host galaxy relations depends on \mbh\ in each simulation. 
We divide each dataset into high- and low-mass groups based on the median \mbh\ value of the sample of the most massive 3607 galaxies within each simulation (the horizontal dashed lines in  Figure~\ref{fig:scale_rels} and the vertical lines in Figure~\ref{fig:MBH_dependence}).
These cutoffs are at $2.6\times10^7 M_\odot$, $1.1\times10^8 M_\odot$, and $1.3\times10^7 M_\odot$ for Illustris, TNG, and EAGLE, respectively.
Most of the observed galaxies with dynamical \mbh\ measurements in \citet{Kormendy_2013} have \mbh\ above these values. 
Additionally, we included a set with only half of the galaxies that are randomly selected and contain the entire range of \mbh\ values (labeled ``Half" in Figure~\ref{fig:MBH_dependence}), in order to achieve a fair comparison with the same sample size as the high- and low-mass groups.

Figure~\ref{fig:MBH_dependence} shows that across all simulations, the low-mass group exhibits more scatter across the feature set compared to the high-mass group. This intrinsically larger scatter can be partially explained by the tear-drop shape that is seen in multiple relations (see Figure~\ref{fig:scale_rels}). The larger scatter at low \mbh\ reduces a regressor's ability to accurately predict \mbh, as opposed to a tighter correlation at high \mbh. As a result, most galaxy properties are only very slightly better correlated with \mbh\ than random numbers for low \mbh\ hosts. Observational, the SMBH--host galaxy relations are also stronger for more massive systems and weaker in smaller galaxies \cite{Graham_2016, Timothy_and_Philip_2014}. For example, \citet{Reines_2015} showed that broad-line AGNs with lower estimated \mbh\ show larger scatters in their BH--host galaxy scaling relations. It is also widely accepted that SMBH feedback has a stronger influence on more massive host galaxies \cite{Croton2006, Fabian2012, Hlavacek-Larrondo2022}. Previous theoretical studies with simulations have also found a similarly larger scatter at the low mass end \citep{Volonteri_2016}.

The full and half populations show similar MSEs in nearly all relations, suggesting that convergence is reached even with only half of the dataset. 
In Illustris and TNG, the MLP regressor consistently performs better on the full set when compared to either the low-mass or high-mass set. This is because the correlation becomes better defined with a fuller range of values, making it easier for machine learning regressors to predict \mbh. On the other hand, in \hbox{EAGLE}, the MSE of the high-mass set is very close to that of the full set.
Given the higher dynamic range that the full set covers, one would expect the MLP regressor to perform better when trained on this data if both the high- and low-mass sets share similar characteristics. However, our result suggests that only Illustris and TNG follow this trend, while EAGLE's low-mass set does not appear to contain any MLP-learnable correlation.
Instead, the \hbox{EAGLE} low-mass galaxies included in the set can act like noise and ``contaminate" the data, which leads to even higher MSE in certain relations (e.g. see the \mbh--\mhalo\ relation in Figure~\ref{fig:MBH_dependence}).

\subsection{Multi-dimensional Correlations}\label{sec:multi}

\begin{figure*}[ht]
\centering
\includegraphics[width=0.99\textwidth]{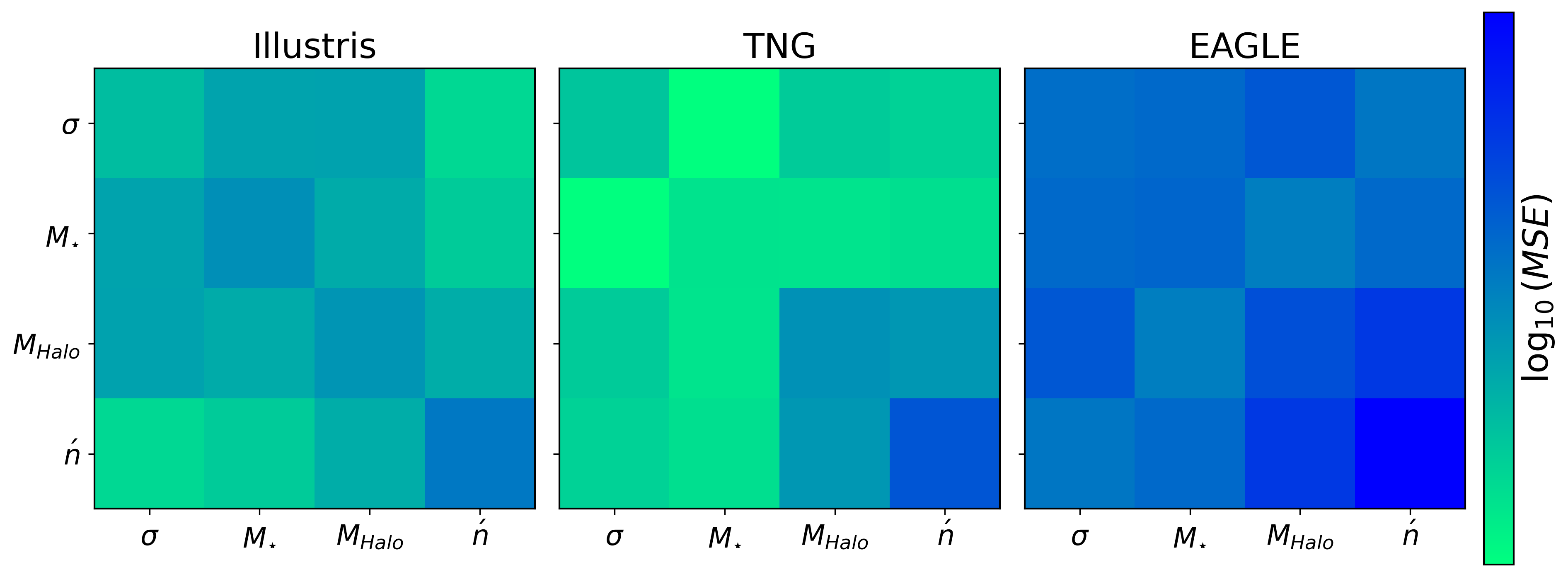}\caption{MLP regressor's measurement of SMBH-host relations when two galaxy properties are combined. The diagonal cells use only one galaxy property (same as the black line in Figure~\ref{fig:MSE_results}).}
\label{fig:Multi-relation}
\end{figure*}

\begin{figure*}[ht]
\centering
\includegraphics[width=0.99\textwidth]{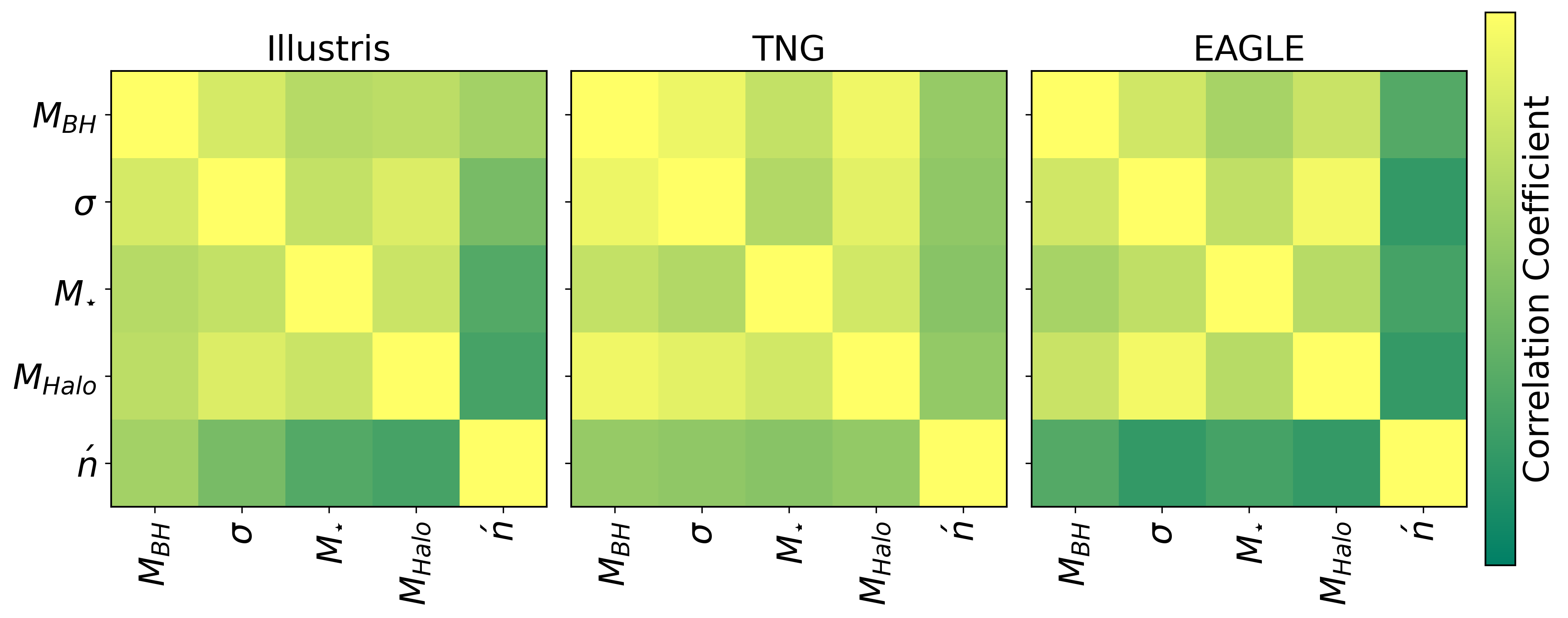}\caption{Spearman correlation matrix of all galaxy properties studied here including \mbh.}
\label{fig:Correlation}
\end{figure*}

In this section, we investigate the ``fundamental plane" of SMBH scaling relations by analyzing the combined effect of two galaxy properties on \mbh\ prediction accuracy. 
In effect, we supply the MLP regressor with a set of two host galaxy properties at a time and evaluate their performance via the MSE reported in Figure~\ref{fig:Multi-relation}.
Additionally, we have included Figure~\ref{fig:Correlation}, which shows the Spearman Rank correlation coefficient of pairs of relations.

In Illustris, the single-property \msig\ relation performs well, but the \mbh--\{\sigmasign,\sersic\} relation (the relation between \mbh\ and a combination of \sigmasign\ and \sersic) yields the most significant improvement in MSE, having a value of 0.0033 compared to 0.0042 of the \mbh--\sigmasign\ relation.
Even though the \mbh--\sersic\ relation performs poorly in isolation, the \mbh--\{\sigmasign,\sersic\} relation complement each other and boost the capability of the MLP in predicting \mbh.
Figure~\ref{fig:Correlation} shows that \sigmasign\ correlates more poorly with \sersic\ than it does with the other galaxy properties, suggesting that \sigmasign\ and \sersic\ are more orthogonal features and are complementary to each other. The enhanced MSE in this case suggests that incorporating galaxy properties with poor correlation with each other can be beneficial, provided at least one of them holds a strong individual correlation with \mbh. On the other hand, when two well-correlated galaxy properties are used, the predicting power may be weakened if one of them is poorly correlated with \mbh (and thus acting as noise/contamination). For example, when combining \sigmasign\ with \mstar or \mhalo, MLP's prediction becomes worse than using \sigmasign\ alone.

In TNG, the \mbh--\{\sigmasign,\mstar\} combination provides the best MSE.
We again see the combination of properties that weakly correlate with each other (\mstar\ and \sigmasign) yet both strongly correlated with \mbh, significantly improves the model’s performance. 
Additionally, since both \sigmasign\ and \mstar\ already strongly correlate with \mbh\ by themselves, it is not a surprise that the combination between these features performs the best out of all combinations.

In EAGLE, similarly to Illustris, the \mbh--\{\sigmasign,\sersic\} relation performs well. The strongest combination is \{\mstar,\mhalo\}. However, the generally larger SMBH-host scatter inherent to the EAGLE simulations limits the MLP's overall ability to correctly predict \mbh even when combining galaxy properties. 

These results underscore the complex coupling between the host galaxy and the SMBH evolution. When two host galaxy properties correlate well with \mbh\ individually, but do not cross-correlate with each other, the combined relation is able to predict \mbh\ better. 
This reveals that machine learning models can capture nuanced interdependencies between galaxy characteristics that traditional analytical methods might miss. 

\section{Conclusions} \label{sec:conclusion}

In this paper, we explore the correlations between SMBHs and their host galaxies in three cosmological simulations: Illustris, TNG, and EAGLE. 
The galaxy properties studies here include stellar velocity dispersion (\sigmasign), dark matter halo mass (\mhalo), stellar mass (\mstar), and the Sersic index (\sersic).  We employ a multitude of machine learning regression techniques to better handle the non-linearity of SMBH-host relations, as well as the multi-dimensional connections between SMBHs and host galaxy properties. 

First, we compare the performance of several regressors(linear regression, decision tree, random forest, multi-layer perceptron (MLP), and XGBoost) on individual SMBH-host relations. Then, taking the best-performing model (the MLP), we carry out further studies on SMBH mass dependence and multi-dimensional correlations. Key findings of our research are as follows:

\begin{enumerate}

    \item For all individual SMBH-host relations in all three simulations, MLP regressor routinely performs the best among all regressors in its ability to correctly predict SMBH masses (based on the SMBH-host relation it has ``learned''). Random forest and XGBoost have similar levels of performance and are both superior to decision tree, while all three perform better than linear regression in handling SMBH-host relations that are fundamentally non-linear, such as the \mbh\-\mhalo relation in Illustris and EAGLE. 

    \item There are some consistent trends in SMBH-host relations across different simulations. The \mbh-\sigmasign relation is a strong correlation in all simulations, showing \sigmasign to be a good predictor of \mbh. However, in TNG, the \mbh-\mstar relation is even tighter than \mbh-\sigmasign, and is the strongest correlation of all studies in this work. \mbh-\sersic\ is relatively weak correlation in all simulations.
    
    \item Different simulations can display distinct SMBH-host relations due to variations in their sub-grid models. For example, the EAGLE simulation produces generally weaker SMBH-host relations. This may be partly due to their delayed SMBH feedback model, which may weaken the intrinsic SMBH-host correlations. EAGLE also has a wider distribution in black hole masses, posing additional challenges for predicting SMBH masses compared to simulations like TNG and Illustris. 

    \item The three simulations produce very different \mbh\ distributions among their most massive 3607 galaxies studied here. EAGLE's SMBHs are the least massive, while TNG's SMBHs are the most massive, with their mean value almost an order of magnitude higher than EAGLE's. Within each simulation, the low mass SMBHs are much more poorly correlated with their host galaxies compared with their high mass counterparts. Still, MLP's ability to predict \mbh in Illustris and TNG improves when all SMBHs are included because of the higher dynamical range.

    \item When combining all the galaxy properties studied here, MLP's ability to predict \mbh\ becomes stronger than any individual SMBH-host relation. This is true in all three simulations, and for both low- and high-mass SMBHs. 
    However, combining only two relations does not always mean improving \mbh\ predictions. Instead, we find that certain pairs of host galaxy properties possess markedly superior predictive power compared to others. In Illustris, this is a combination of \sigmasign\ and \sersic. Most likely, these galaxy properties are \textit{orthogonal} to each other, that is, they are minimally cross-correlated, while both are strongly correlated with \mbh\ in isolation. Training the MLP with both \sigmasign\ and \sersic\ then provides the model with a highly detailed set of relations to learn from. In TNG, the strongest combination is \sigmasign\ and \mstar.
    
\end{enumerate}

Our work shows that cosmological simulations with different subgrid models produce different relations between SMBHs and their host galaxy properties. Machine learning offers a powerful approach for measuring these SMBH-host relations, particularly when they are inherently non-linear or involve higher-dimensional correlations. Unlike traditional methods, machine learning regressors provide a flexible way to quantify the tightness of scaling relations without assuming a specific functional form, thereby minimizing biases. In future work, we aim to expand our analysis to higher redshifts and to include additional simulations and scaling relations. As larger observational datasets become available, applying this framework to compare observations with simulations will yield valuable insights into the coevolution of SMBHs and galaxies.

\begin{acknowledgements}

We would like to thank Francisco Villaescusa-Navarro, Min Long, and Marco Buongiorno Nardelli for helpful discussions. Y.L. acknowledges financial support from NSF grants AST-2107735 and AST-2219686, NASA grant 80NSSC22K0668, and Chandra X-ray Observatory grant TM3-24005X. 

\end{acknowledgements}

\appendix

\section{Hyperparameter Optimization} 


Our hyperparameter optimization is conducted using the sklearn RandomizedSearchCV function. We use cross validation, with a shuffling Kfold of 5. This helps combat overfitting and keep the results of all regressors comparable, regardless of model complexity. We applied this method to all regressors used in this work, where each had a list of hyperparameters and values that were tested. Table~\ref{tab:tested_parameters} lists the hyperparameters tested and optimized for the MLP regressor, which we use as a representative regressor here. The optimized hyperparameters to produce the lowest MSE for all relations in Figure~\ref{fig:MSE_results} using the MLP regressor are listed in Table~\ref{tab:combined-mlp-params}.

\begin{deluxetable}{cc} 
\tablecaption{Tested Hyperparameters for MLP\label{tab:tested_parameters}}
\tablehead{\colhead{Hyperparameter Name} & \colhead{Values Tested}}
\startdata
Number of Layers & [1,2,3]\\
Hidden Layer Size & [50,100,250,500,750,1000]\\
Activation Function & [relu, tanh, logistic]\\
Solver & [adam, sgd, lbfgs]\\
Alpha & [$10^{-6}$, $10^{-5}$, $10^{-4}$, $10^{-3}$]\\
Learning Rate & [constant, invscaling, adaptive]\\
Learning Rate Initial & [$10^{-6}$, $10^{-3}$,$10^{-1}$]\\
Maximum Iterations & [200, 300, 500]\\
Tolerance & [$10^{-4}$,$10^{-3}$]\\
Early Stopping & [True, False]\\
\enddata
\end{deluxetable}


\begin{longtable}[ht]{llcccccc}
\caption{MLP Model Parameters Across Simulations and Galaxy Parameters} \label{tab:combined-mlp-params} \\
\toprule
Simulation & Hyperparameter & All & $\sigma$ & $M_*$ & $M_{\mathrm{halo}}$ & $n_{\mathrm{sersic}}$ & Random \\
\midrule
\endfirsthead

\multicolumn{8}{c}{{\bfseries Table \thetable\ continued from previous page}} \\
\toprule
Simulation & Parameter & All & $\sigma$ & $M_*$ & $M_{\mathrm{halo}}$ & $n_{\mathrm{sersic}}$ & Random \\
\midrule
\endhead

\bottomrule
\endfoot

\bottomrule
\endlastfoot

\multirow{9}{*}{Illustris}
 & Tolerance & 0.0001 & 0.0001 & 0.0001 & 0.0001 & 0.0001 & 0.001 \\
 & Solver & lbfgs & lbfgs & lbfgs & lbfgs & lbfgs & adam \\
 & Maximum Iterations & 500 & 500 & 200 & 200 & 500 & 300 \\
 & Initial Learning Rate & 0.01 & 1e-05 & 1.0 & 0.01 & 1e-05 & 0.01 \\
 & Learning Rate Type & invscaling & adaptive & adaptive & constant & invscaling & invscaling \\
 & Hidden Layers & 100 & (100,100,100) & (50,50,50) & (250,250) & (1000,1000,1000) & 50 \\
 & Early Stopping & False & False & True & False & False & True \\
 & Alpha & 0.001 & 0.01 & 0.01 & 0.01 & 0.01 & 0.01 \\
 & Activation Function & relu & relu & relu & relu & relu & relu \\

\midrule
\multirow{9}{*}{TNG}
 & Tolerance & 0.0001 & 0.0001 & 0.0001 & 0.0001 & 0.0001 & 0.001 \\
 & Solver & adam & adam & adam & lbfgs & lbfgs & sgd \\
 & Maximum Iterations & 300 & 300 & 500 & 300 & 500 & 300 \\
 & Initial Learning Rate & 0.01 & 0.01 & 0.01 & 0.01 & 1.0 & 1.0 \\
 & Learning Rate Type & constant & adaptive & adaptive & constant & adaptive & invscaling \\
 & Hidden Layers & (250,250,250) & (1000,1000,1000) & (1000,1000,1000) & (250,250) & (250,250) & 50 \\
 & Early Stopping & True & True & True & False & True & True \\
 & Alpha & 0.0001 & 0.0001 & 1e-05 & 0.001 & 0.001 & 0.001 \\
 & Activation Function & relu & relu & relu & relu & relu & relu \\

\midrule
\multirow{9}{*}{Eagle}
 & Tolerance & 0.0001 & 0.0001 & 0.0001 & 0.0001 & 0.0001 & 0.001 \\
 & Solver & lbfgs & lbfgs & lbfgs & lbfgs & lbfgs & adam \\
 & Maximum Iterations & 500 & 300 & 200 & 500 & 500 & 500 \\
 & Initial Learning Rate & 0.01 & 1e-05 & 1e-05 & 1.0 & 1.0 & 0.01 \\
 & Learning Rate Type & constant & constant & invscaling & adaptive & constant & invscaling \\
 & Hidden Layers & (500,500,500) & (1000,1000) & 100 & (50,50,50) & (100,100,100) & (500,500) \\
 & Early Stopping & False & False & False & True & True & True \\
 & Alpha & 1e-05 & 0.01 & 0.0001 & 0.001 & 0.001 & 0.0001 \\
 & Activation Function & relu & relu & relu & relu & relu & relu \\

\end{longtable}

\end{document}